\documentclass[sigconf,authorversion,nonacm]{acmart}

\AtBeginDocument{%
  \providecommand\BibTeX{{%
    \normalfont B\kern-0.5em{\scshape i\kern-0.25em b}\kern-0.8em\TeX}}}

\begin{document}

\title{ReSiPI: A Reconfigurable Silicon-Photonic 2.5D Chiplet Network with PCMs for Energy-Efficient Interposer Communication\vspace{-0.05in}}


\author{Ebadollah Taheri, Sudeep Pasricha, and Mahdi Nikdast}
\affiliation{%
  \institution{Department of Electrical and Computer Engineering, Colorado State University, USA\vspace{-0.11in}}
  \country{}
}

\begin{abstract}
2.5D chiplet systems have been proposed to improve the low manufacturing yield of large-scale chips. However, connecting the chiplets through an electronic interposer imposes a high traffic load on the interposer network. Silicon photonics technology has shown great promise towards handling a high volume of traffic with low latency in intra-chip network-on-chip (NoC) fabrics. Although recent advances in silicon photonic devices have extended photonic NoCs to enable high bandwidth communication in 2.5D chiplet systems, such interposer-based photonic networks still suffer from high power consumption. In this work, we design and analyze a novel \underline{Re}configurable power-efficient and congestion-aware \underline{Si}licon-\underline{P}hotonic 2.5D \underline{I}nterposer network, called ReSiPI. Considering runtime traffic, ReSiPI is able to dynamically deploy inter-chiplet photonic gateways to improve the overall network congestion. ReSiPI also employs switching elements based on phase change materials (PCMs) to dynamically reconfigure and power-gate the photonic interposer network, thereby improving the network power efficiency. Compared to the best prior state-of-the-art 2.5D photonic network, ReSiPI demonstrates, on average, 37\% lower latency, 25\% power reduction, and 53\% energy  minimization in the network. \vspace{-0.1in}
\end{abstract}



\keywords{NoC, 2.5D chiplet system, silicon photonics, interposer, PCM.\vspace{-0.1in}}


\maketitle

\section{Introduction}
\begin{figure}[t]
\centering
\includegraphics[scale=0.40]{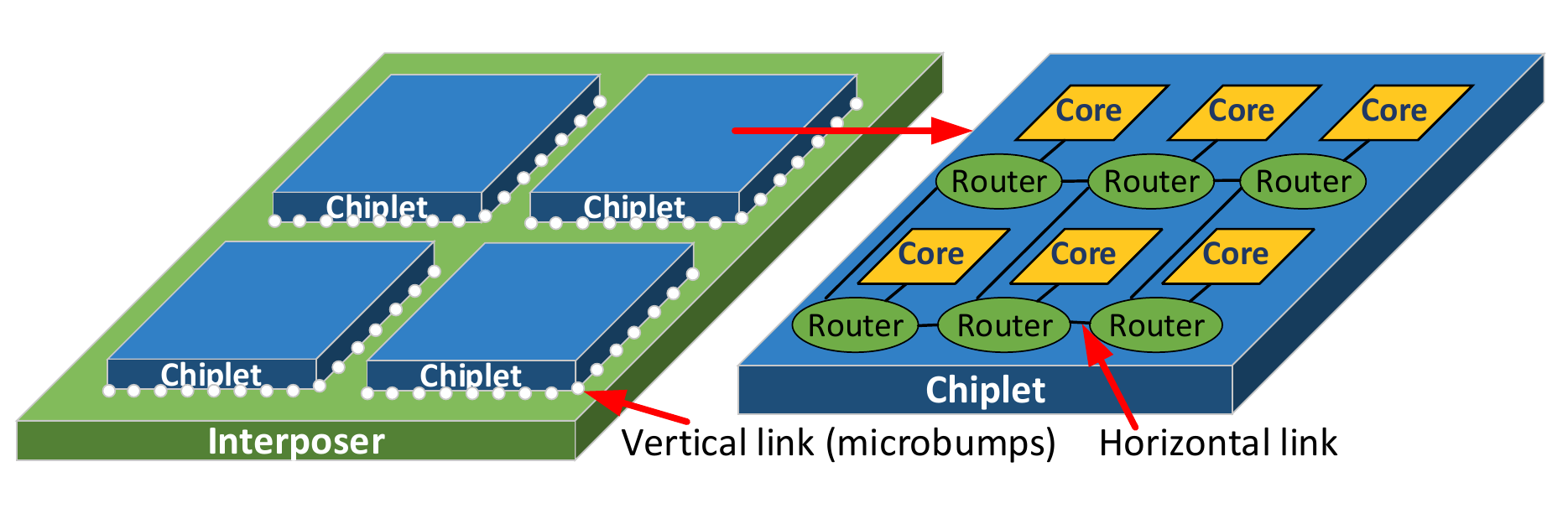}
\vspace{-0.25in}
\caption{An example of a 2.5D chiplet system with four chiplets connected through an interposer. }\label{Fig:IntroSec:ChipletSystem}
\vspace{-0.25in}
\end{figure}
The continuous growth in data- and compute-intensive applications (e.g., big data analytics and deep learning) necessitates the design of large-scale chips with high compute performance and with a large degree of parallelism. Such large-scale chips include many processing cores, ranging from a few tens to hundreds~\cite{davies2018loihi}. Although such manycore chips offer the computational capabilities required to support emerging applications, they suffer from a low manufacturing yield due to their large chip size~\cite{kannan2015enabling}. To address this problem, 2.5D chiplet systems have been introduced in which a large chip is disintegrated into several smaller chips, called chiplets, that are connected through an inter-chiplet interposer network, to significantly improve the collective manufacturing yield~\cite{kannan2015enabling, bharadwaj2020kite}. 

An example of a 2.5 chiplet system is shown in Fig.~\ref{Fig:IntroSec:ChipletSystem}. In this example, four chiplets are integrated on an interposer---with 2.5D technology---which enables inter-chiplet communication~\cite{taheri2021deft, majumder2020remote}. Moreover, each chiplet includes six processing cores interconnected with an intra-chiplet mesh-based network-on-chip (NoC). While such a 2.5D chiplet system provides higher modularity and yield than a monolithic 2D chip with the same functionality, the interposer network becomes a potential bottleneck as it is supposed to provide low-latency and high bisection-bandwidth communication among the chiplets, which significantly impacts the system's performance and scalability. Although conventional electronic NoCs can efficiently support a small chip with low to medium traffic load, such as at the intra-chiplet level, they impose a high latency when they are employed on an interposer to handle the global traffic among chiplets~\cite{thonnart2020popstar, narayan2020prowaves, fotouhi2019enabling}. The high latency of an electronic interposer is due to its long metal interconnects and low inherent bandwidth to support the high volume inter-chiplet traffic.

To improve intra-chip communication performance in manycore systems, photonic NoCs (PNoCs) \cite{sunny2021arxon, mirza2020opportunities, chittamuru2018bignoc, chittamuru2017swiftnoc}, which use silicon photonic devices and waveguides to modulate, switch, and transmit data among many processing cores and memory, can be used. Advances in silicon photonics technology \cite{werner2017survey} have allowed data transmission in PNoCs to benefit from the high throughput, reduced dynamic power, and lower transmission delays of light-speed communication~\cite{pasricha2020survey}. The inherent high bandwidth and low latency of PNoCs also makes them a promising solution for inter-chiplet communication in 2.5D platforms~\cite{narayan2020prowaves, thonnart2020popstar}. Accordingly, 2.5D chiplet systems with photonic interposer networks have recently received some attention~\cite{fotouhi2019enabling, thonnart2020popstar, narayan2020prowaves, zheng2020versatile}. Such photonic interposer networks can employ wavelength-division multiplexing (WDM) to simultaneously support multiple data streams, each modulated on a different optical wavelength traversing a waveguide, to boost communication bandwidth. However, a high bandwidth photonic interposer network also requires a large number of wavelengths per waveguide, which imposes a high laser power consumption overhead~\cite{narayan2020prowaves}. Fortunately, as we show in this work, a reconfigurable photonic interposer network can handle such power-performance trade-off, where the network bandwidth can be increased for high traffic load scenarios, and similarly, the bandwidth can be reduced to save power under low traffic load conditions.

More recently, the integration of silicon photonics and phase-change materials (PCMs) has created a unique opportunity to realize adaptable, reconfigurable, and programmable photonic networks. PCM-based switches~\cite{zhang2020ultra, xu2019low} and couplers~\cite{teo2022comparison} have been proposed to realize energy-efficient optical signal switching in photonic networks. In particular, PCM-based silicon photonic devices are non-volatile devices in which the switching state is preserved even in the absence of an electrical voltage/current, hence improving the power efficiency in networks employing such devices. Although PCM-based devices are relatively slow (e.g., 10~Mhz~\cite{zhang2020ultra}) to be used for fast switching, they are still very efficient devices to support sporadic network reconfiguration \cite{zolfaghari2022non}.        

Prior work has explored the use of silicon photonics to realize high performance interposer networks \cite{fotouhi2019enabling, thonnart2020popstar, narayan2020prowaves, zheng2020versatile}. However, these efforts either suffer from high power consumption or high network latency because the interposer is not reconfigurable to adapt to different traffic load conditions. To address these drawbacks, we develop a novel PCM-based \underline{Re}configurable \underline{Si}licon-\underline{P}hotonic \underline{I}nterposer (ReSiPI) network for 2.5D chiplet systems. The main contributions of ReSiPI are summarized below.
\begin {itemize}
\item We propose a reconfigurable photonic interposer network with an intelligent dynamic gateway-activation mechanism based on the network's traffic load at runtime.
\item ReSiPI increases inter-chiplet communication bandwidth by increasing the number of active gateways, and not wavelengths, to efficiently distribute the bandwidth improvement across chiplets while saving laser power. 
\item We present a power-saving mechanism to tune input optical power of modulators by employing PCM-based devices and laser-power management in ReSiPI.
\item As source routers on chiplets need to select a gateway to send packets to other chiplets, ReSiPI proposes an efficient dynamic gateway-selection approach to distribute traffic load while minimizing source-destination hop-counts.  
\end{itemize}

The rest of the paper is organized as follows. Background and prior related work in 2.5D chiplet systems are reviewed in Section 2. Section 3 discusses our proposed photonic interposer network, ReSiPI. In Section 4, evaluation results comparing ReSiPI to the state-of-the-art are presented. Finally, Section 5 concludes the paper.

\section{Background and Related Work}

\subsection{Chiplet systems and electronic interposers}
To improve manufacturing costs, 2.5D integration was employed in \cite{kannan2015enabling} to disintegrate a large multicore chip into smaller chiplets. Doing so breaks the original, larger NoC into several smaller NoCs on each chiplet and an inter-chiplet interposer network. However, such a disintegration introduces some performance loss in the system because it is not trivial to create an interposer network that can support high bandwidth and fast communication required among chiplets. Moreover, the disintegration of the original deadlock-free NoC can introduce new system-wide deadlock conditions where a cyclic dependency of requests for buffer resources among different chiplets and the interposer negatively affects the system performance. To address deadlock, \cite{yin2018modular} and \cite{majumder2020remote} proposed routing algorithms to avoid deadlock in 2.5D chiplet systems.

In addition to deadlock, the interposer network can suffer from traffic congestion especially when the system scales up~\cite{taheri2021deft}. As shown in Fig.~\ref{Fig:IntroSec:ChipletSystem}, there are multiple chiplets and each with several integrated cores, all of which communicate through the interposer network. Therefore, the interposer network should be able to handle a high volume of traffic among chiplets. Moreover, the interposer is large and metal interconnects impose a high delay for long-distance communication \cite{mekawey2022optical}. To this end, silicon-photonic interposers have been proposed to improve the latency and bandwidth compared to conventional electronic interposer networks~\cite{thonnart2020popstar, narayan2020prowaves, fotouhi2019enabling}. 

\begin{figure}[t]
\centering
\includegraphics[scale=0.445]{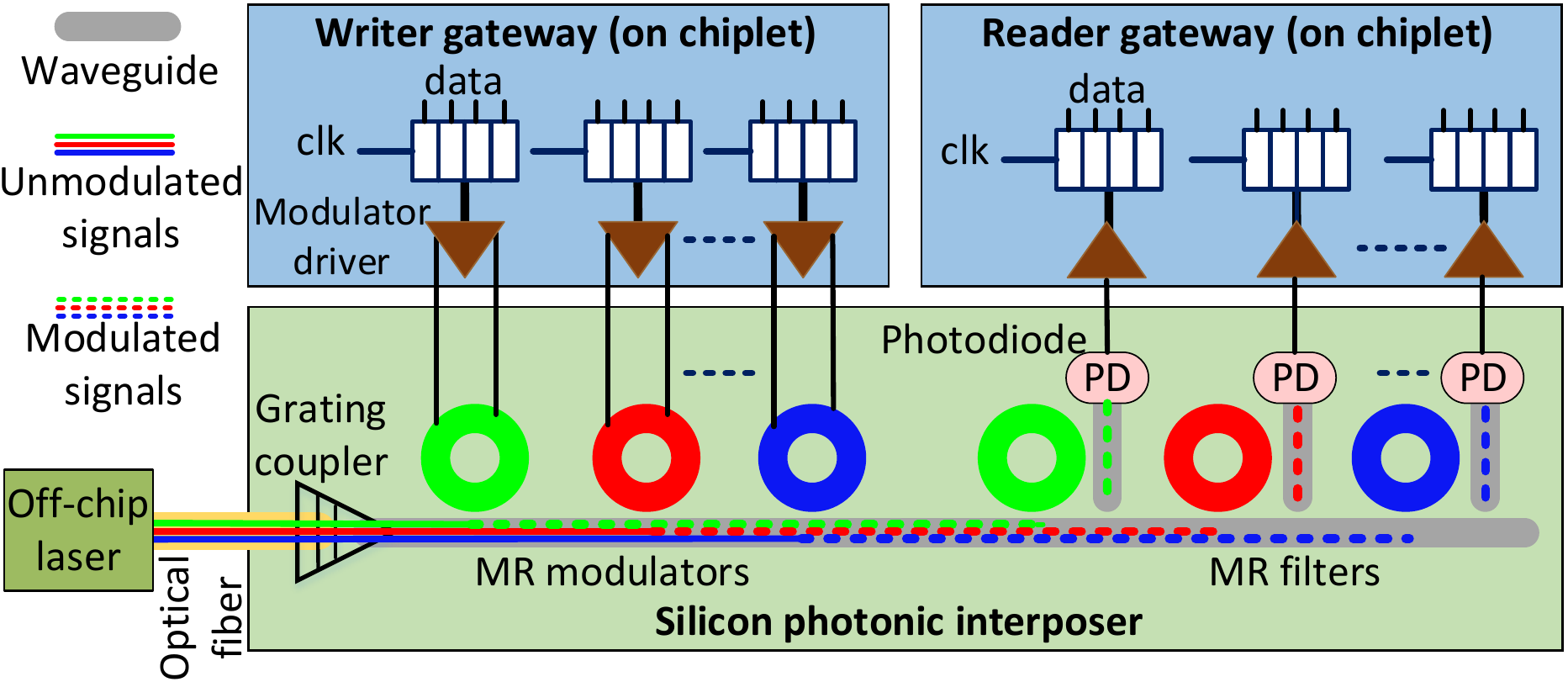}
\vspace{-0.3in}
\caption{Data transmission on a photonic interposer.}
\vspace{-0.25in}
\label{Fig:BackgroundSec:OpticLink}
\end{figure}

\subsection{Silicon photonic interposers}
 Fig.~\ref{Fig:BackgroundSec:OpticLink} shows an example of data transmission between two chiplets that are placed on a silicon-photonic interposer. On the interposer, some modulators, filters, and photodiodes (PDs) are used to perform electro-optical and opto-electrical data conversions. We consider microring resonator (MR) devices~\cite{mirza2021silicon, mirza2020silicon} for modulators and filters, due to their area and power efficiency. We define a gateway as an electronic circuit on a chiplet, which controls the modulators (i.e., writer gateway in Fig.~\ref{Fig:BackgroundSec:OpticLink}) and PDs (i.e., reader gateway in Fig.~\ref{Fig:BackgroundSec:OpticLink}) on the interposer. Moreover, gateways receive/send data from/to the routers on the same chiplet. As shown in Fig.~\ref{Fig:BackgroundSec:OpticLink}, optical signals with different wavelengths are generated in an off-chip laser source (green, red, and blue wavelengths). The optical signals are then coupled to the waveguide on the photonic interposer using an optical fiber and a grating coupler. At the writer gateway, MRs modulate electronic data on the optical signals and, at the reader gateway, each MR filters its corresponding optical signal to be detected by the PD. Note that each MR device is designed to resonate at (i.e., couple with) a specific wavelength. As a result, several wavelengths can transmit bits of data at the same time, over the same waveguide; this technique is called WDM (see Section 1). 

Employing silicon photonics, \cite{fotouhi2019enabling} proposed an interposer based on arrayed-waveguide grating routers (AWGRs) to improve the high latency of electronic interposers. However, \cite{fotouhi2019enabling} considered static optical bandwidth under different traffic loads, which either wastes system power under low traffic loads or sacrifices performance under high traffic loads. PROWAVES in~\cite{narayan2020prowaves} proposed a dynamic bandwidth-management technique for optical gateways by adjusting the number of active wavelengths with respect to the runtime traffic load. The number of active wavelengths is updated in a time epoch based on the network delay experienced in the previous epochs. However, using a single high bandwidth gateway to support several routers on a chiplet creates contention among the intra-chiplet routers to access the high-bandwidth gateway. As we will discuss, ReSiPI increases the number of active gateways to improve optical bandwidth while, at the same time, the gateways are distributed over the chiplet to improve router-gateway access and network congestion. Moreover, ReSiPI intelligently power-gates the idle gateways and manages the input laser power based on the runtime traffic load, to improve the interposer energy-efficiency.

\subsection{PCM-based silicon photonic devices}
Photonic devices based on phase-change materials (PCMs) have recently received attention due to their non-volatile property which helps save static tuning power consumption in photonic-switched networks~\cite{teo2022comparison, zhang2020ultra, xu2019low, wuttig2017phase}. A PCM has two states with different optical properties: amorphous and crystalline states. A short optical or electrical pulse can switch the states~\cite{wuttig2017phase} while a state can be preserved without consuming any power. As the amorphous and crystalline states have different optical properties, PCM-based devices are attractive to design non-volatile optical switches and couplers for photonic networks. For example, a broadband PCM-based switch was proposed in \cite{xu2019low} which requires $\approx$2~nJ energy for reconfiguration. In~\cite{zolfaghari2022non}, the same switch was employed to power-gate MR filters in idle reader nodes to reduce a PNoC's tuning power consumption. However, the proposed architecture in \cite{zolfaghari2022non} does not account for dynamic bandwidth management in the network, to handle the runtime traffic. Moreover, the main power consumption in PNoCs comes from the laser source \cite{pasricha2020survey}, while \cite{zolfaghari2022non} only accounts for MR tuning power consumption.  

\section{R\lowercase{e}S\lowercase{i}PI: Overview}
\begin{figure}[t]
\centering
\includegraphics[scale=0.34]{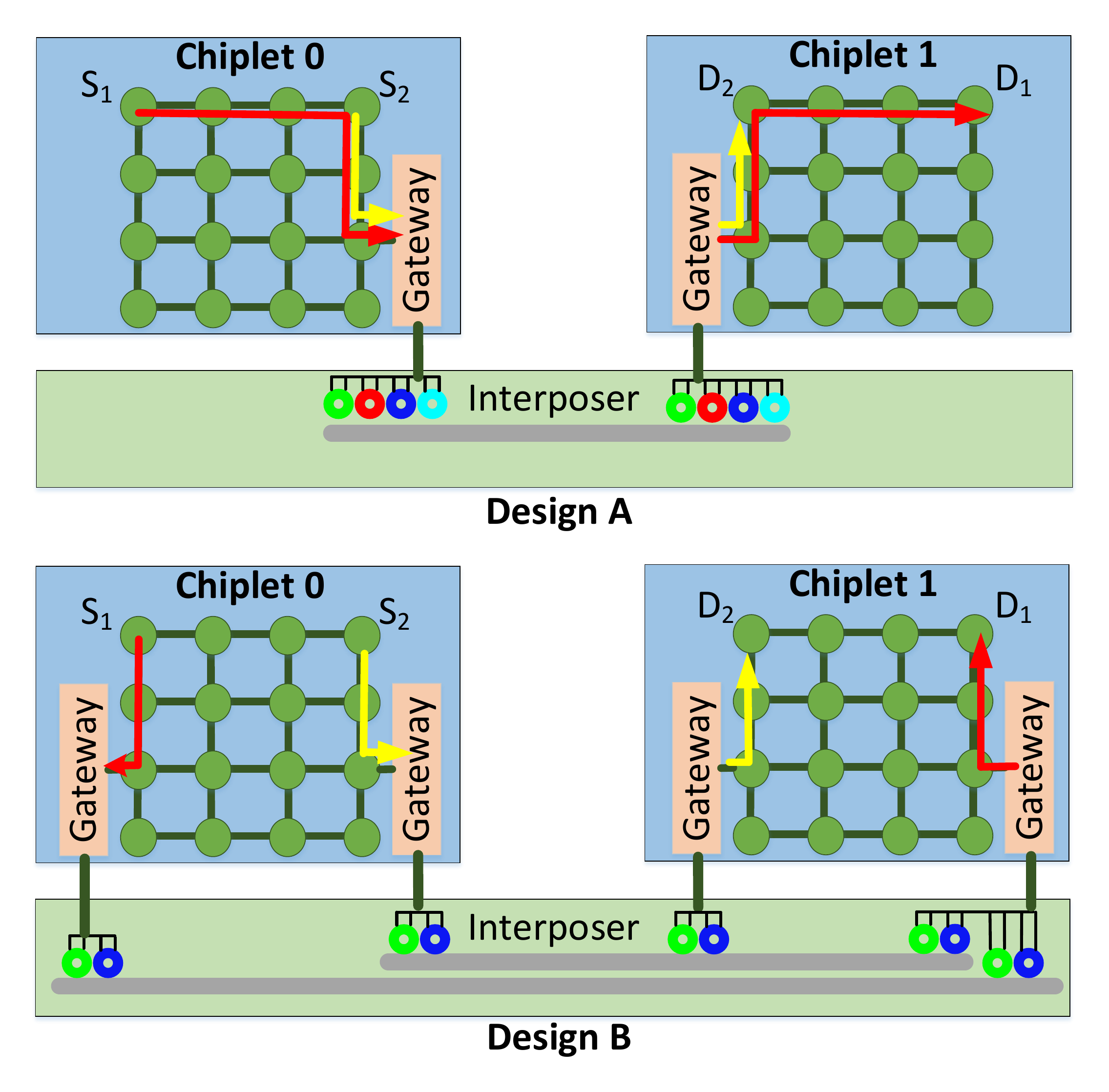}
\vspace{-0.17in}
\caption{Dynamic inter-chiplet bandwidth management: Design A with a larger number of wavelengths (e.g., four) and Design B, which is considered in ReSiPI, with a larger number of gateways and fewer (e.g., two) wavelengths.}
\label{Fig:PropSec:Motivation}
\vspace{-0.23in}
\end{figure}

\begin{figure}[t]
\centering
\includegraphics[scale=0.28]{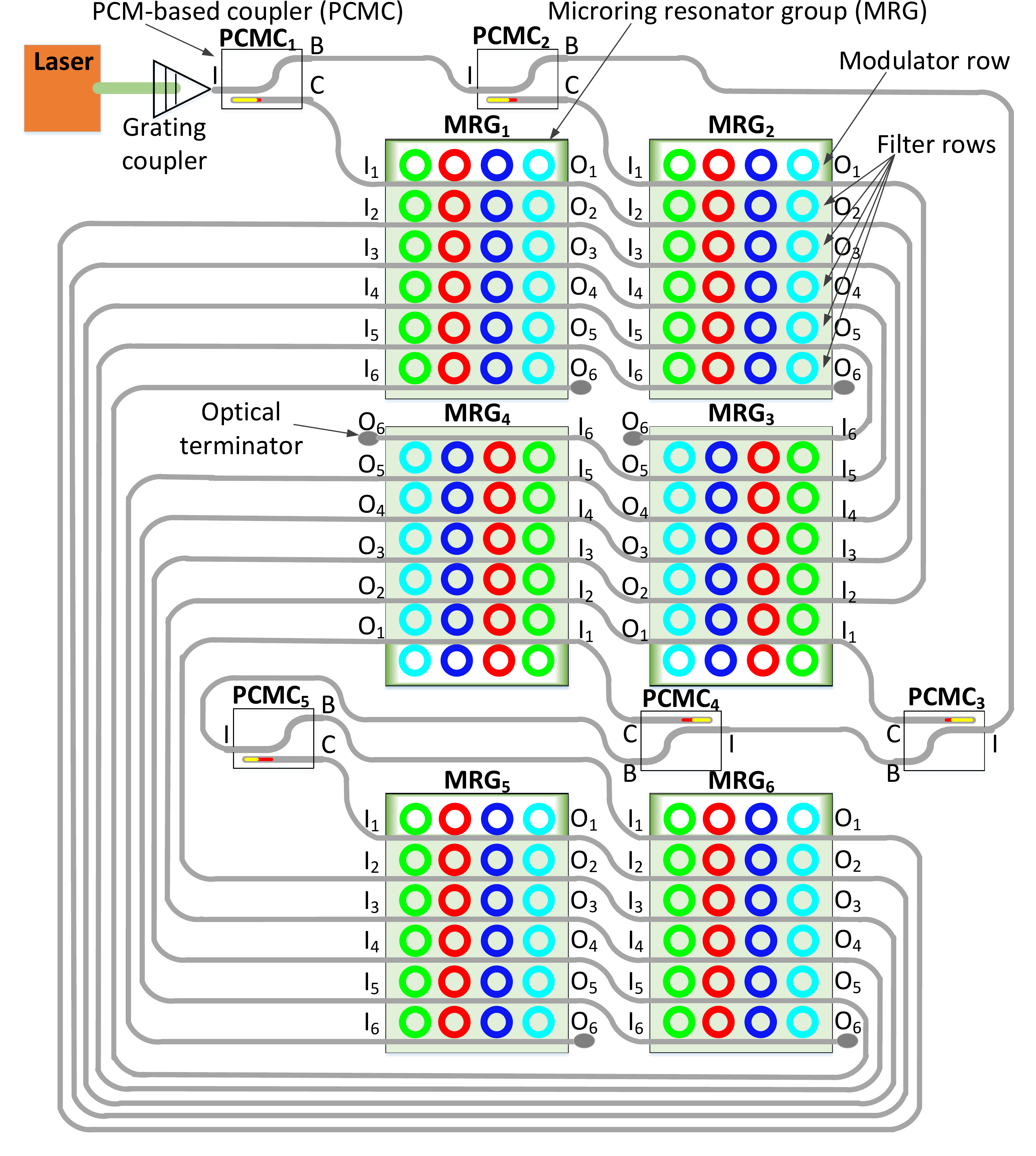}
\vspace{-0.3in}
\caption{An example of the proposed photonic interposer architecture (ReSiPI) with a total of six gateways (one per chiplet) and four optical wavelengths. This architecture can be extended to have multiple gateways per chiplet.}
\label{Fig:PropSec:ReSiPIInterposer}
\vspace{-0.2in}
\end{figure}

In our ReSiPI architecture, an electronic intra-chiplet NoC is considered on each chiplet and a silicon photonic network is considered for the inter-chiplet interposer network. ReSiPI employs the gateway configuration in~\cite{thonnart2020popstar, narayan2020prowaves}, where gateways to the interposer are placed on the chiplets. The photonic devices are placed on the interposer, and microbump vertical links are used to pass control signals from the gateway (e.g., for driving modulators) to the silicon photonic devices on the interposer. In this section, after motivating dynamic gateway management, we describe the ReSiPI architecture and its fundamental operational mechanisms.

\subsection{Dynamic gateway management}
Unlike state-of-the-art photonic interposer networks \cite{narayan2020prowaves, narayan2019waves}, where inter-chiplet bandwidth is increased by utilizing a large number of wavelengths, ReSiPI manages the bandwidth by dynamically adjusting the number of active gateways on each chiplet (we consider four gateways per chiplet in our evaluation in Section \ref{Sec:Results}). Fig.~\ref{Fig:PropSec:Motivation} motivates ReSiPI's dynamic gateway-management approach. As can be seen, there are two ways to increase the inter-chiplet bandwidth: 1) using design A with a larger number of active wavelengths (similar to the approach in \cite{narayan2020prowaves}), and 2) using design B with a larger number of active gateways (developed in ReSiPI). In this example, there are two packets going from Chiplet 0 to Chiplet 1: $S_1\rightarrow D_1$ and $S_2\rightarrow D_2$. Let us assume that each of the two packets requires bandwidth proportional to two optical wavelengths. In design A, four wavelengths are activated on the same gateway, while in design B, two gateways with two wavelengths are considered. In design B, as there are two gateways in Chiplet 0, packets can select between two gateways, resulting in a better traffic-load distribution. In other words, not only can the traffic load be better distributed between gateways but also the chiplet's intra-network traffic load can be better distributed across the chiplet's routers. Therefore, design B has the potential to offer higher performance as the intra- and inter-chiplet bandwidth is more efficiently distributed. Moreover, unlike design A, design B can allow for an intelligent gateway selection mechanism to further reduce the source-to-destination hop count. In this example, packet $S_1\rightarrow D_1$ requires ten hops of intra-chiplet routing in design A, while it can be routed with only four hops of intra-chiplet routing in design B. As a result, improving the bandwidth with more number of gateways (as in design B) can result in a better performance-cost trade-off than in the case where number of wavelengths is increased (as in design A).

\subsection{ReSiPI interposer network architecture}
An example of the ReSiPI interposer network with six gateways and four wavelengths is shown in Fig.~\ref{Fig:PropSec:ReSiPIInterposer}. There is a microring resonator group (MRG) associated with each gateway. Each MRG has four columns of MRs with different colors to show the four different optical wavelengths in this example. ReSiPI employs the Single-Writer Multiple-Reader (SWMR) protocol \cite{pasricha2020survey} in waveguides. The first row in each MRG consists of modulator MRs, to actively write electronic data on the associated wavelength on the waveguide (see Fig.~\ref{Fig:BackgroundSec:OpticLink}). The last five rows are filter MRs that are wavelength-selective devices to passively read data from their associated wavelengths on each waveguide. There are five rows of MR filters as each gateway receives data from the other five gateways.
\begin{figure}[t]
\centering
\includegraphics[scale=0.26]{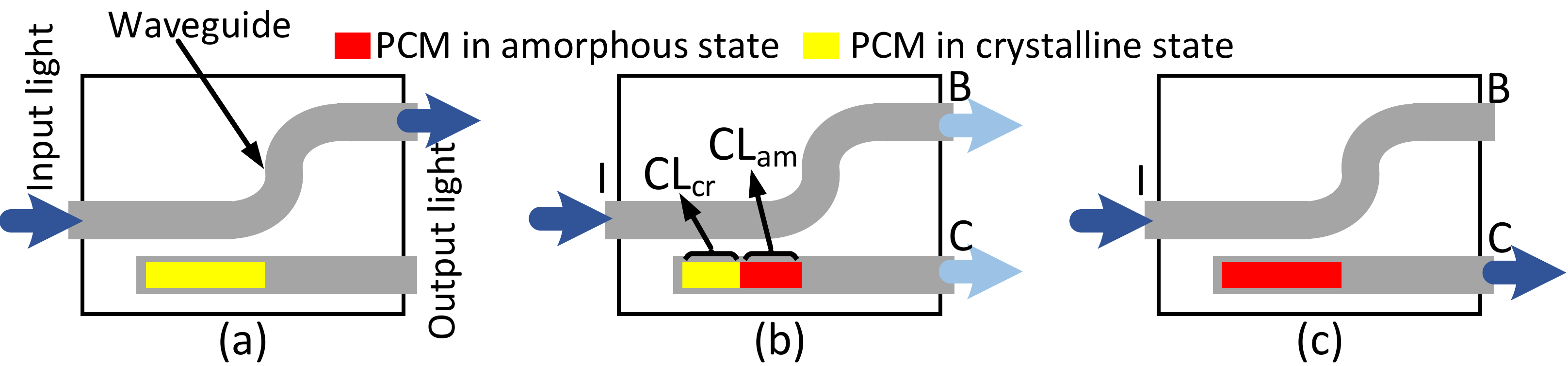}
\vspace{-0.3in}
\caption{A PCM-based coupler (PCMC) in different states: (a) crystalline state to guide light to Bar (B) output, (b) partially crystalline state to guide a portion of light to the Cross (C) output and the rest to the Bar output, and (c) amorphous state to guide the input light to the Cross output.}
\label{Fig:PropSec:Coupler}
\vspace{-0.2in}
\end{figure}

To efficiently mange the power consumption in the network, ReSiPI not only power-gates the idle electronic gateways on the chiplets, but also the power on the optical signals entering the MRGs of idle gateways is appropriately readjusted. To do so, the laser is tuned to generate less optical power at its output and, therefore, consumes less input power. To power-gate the input of MRGs, a non-volatile, PCM-based reconfigurable directional coupler (PCMC)~\cite{teo2022comparison} is employed, as shown in Fig.~\ref{Fig:PropSec:Coupler}. It utilizes PCM to divide the input optical signal between the Cross (C) and Bar (B) outputs. In Fig.~\ref{Fig:PropSec:Coupler}.a, the PCM is completely in the crystalline state and all the input light goes to the B output. In Fig.~\ref{Fig:PropSec:Coupler}.b, where the PCM is partially in the amorphous state, a portion of the input light goes to the C output and the rest traverses to the B output. In Fig.~\ref{Fig:PropSec:Coupler}.c, all the input light propagates to the C output as the PCM is completely in the amorphous state. One practical way to adjust the PCM state is by using an embedded microheater on top of the PCM material on the waveguide~\cite{teo2022comparison}, because the PCM state changes with a temperature change. For example, using a transparent conductive heater, the PCMC can work at the frequency of 10~Mhz~\cite{zhang2020ultra}.

The coupling ratio ($CR$) in the PCMC can be defined as:
\begin{equation}
\kappa=\frac{CL_{am}}{CL_{cr}},
\end{equation}
where $CL_{am}$ and $CL_{cr}$ are the coupling lengths of the amorphous and crystalline states, respectively (see Fig.~\ref{Fig:PropSec:Coupler}.b). By adjusting this coupling ratio (e.g., using a microheater), we can tune the portion of the input light transmitted to the Bar and Cross outputs in a PCMC. Accordingly, and assuming a lossless optical transmission, the optical power at the Cross ($P_{C}$) and Bar ($P_{B}$) output is:
\begin{equation}
P_{C}=\kappa\times P_I,
\end{equation}
\begin{equation}
P_{B}=(1-\kappa)\times P_I,
\end{equation}
where $P_I$ is the input optical power in the PCMC. In our ReSiPI interposer network architecture, the coupling ratio ($\kappa$) is tuned to manage the input laser power on each waveguide.
Considering Fig.~\ref{Fig:PropSec:ReSiPIInterposer}, a PCMC controls the input optical power of each writer. The coupling ratio of PCMCs are tuned based on the total number of active gateways. If the associated writer gateway of PCMC$_i$ is deactivated, the coupling ratio of the PCMC should be zero (i.e., $\kappa_i=$~0, PCM is completely in the crystalline state). Otherwise, the coupling ratio of PCMC$_i$ is:
\begin{equation}
\kappa_i=\frac{1}{(\sum_{c=1}^{C}g_c)-i},
\label{Equ:PropSec:CRActive}
\end{equation}
where $C$ is the total number of chiplets in the system and $g_c$ is the number of active gateways of chiplet $c$.

The organization of MRGs and PCMCs, shown in Fig.~\ref{Fig:PropSec:ReSiPIInterposer}, can be scaled with any number of gateways, chiplets, and PCMCs without loss of generality. Assuming $N$ gateways in the system, the number of MRGs is $N$ while the number of PCMCs is $N-1$. Moreover, the number of the MRs in each MRG is equal to the number of wavelengths and the number of the waveguides in each MRG is $N$. Even rows of MRGs (e.g., the second row with MRG$_3$ and MRG$_4$ in Fig.~\ref{Fig:PropSec:ReSiPIInterposer}) and their PCMCs (e.g., PCMC$_3$ and PCMC$_4$ in Fig.~\ref{Fig:PropSec:ReSiPIInterposer}) are rotated at 180 degrees compared to the odd rows. For any MRG$_k$, if $k<N$, $O_j$ of MRG$_k$ is connected to $I_{j+1}$ of MRG$_{k+1}$ (see MRG connections in Fig.~\ref{Fig:PropSec:ReSiPIInterposer}). When $k=N$, $O_j$ of MRG$_N$ is connected to $I_{j+1}$ of MRG$_1$. Additionally, $I_1$ of MRG$_k$, if $k<N$, is connected to output C of PCMC$_k$. Also, output B of PCMC$_j$, if $j<N-1$, is connected to input I of PCMC$_{j+1}$ (see PCMC connections in Fig.~\ref{Fig:PropSec:ReSiPIInterposer}).

\subsection{Adaptive active gateway selection}
\label{Sec:PropSec:NumOfGates}
ReSiPI aims to assign traffic load to gateways in a manner that minimizes congestion. Nevertheless, if the assigned load is too low---i.e, gateways are underutilized---the system power is wasted. The unnecessary power wastage is due to a larger than needed number of gateways that are activated, and their associated tuning and laser power overhead. Therefore, ReSiPI optimizes the number of active gateways per chiplet with the goal of a trade-off between the overall system power and the average packet latency, in a way that gateways are neither congested nor underutilized. To accomplish this, we define $L_m$ as the maximum allowable load on a gateway. This means that beyond $L_m$ load on a gateway, we can expect congestion and performance loss. We use the maximum packet transmission rate on a gateway to measure the gateway load. Then, we update the number of active gateways in each chiplet based on the average load on each chiplet's gateways, with respect to $L_m$. We discuss how to select an optimal value for $L_m$ in Section~\ref{Sec:Results:DSE}. 

The average gateway load for chiplet $c$ in a reconfiguration interval $i$ ($L^i_c$) is defined as:
\begin{equation}
L^i_c=\frac{1}{g_c}\sum_{j=1}^{g_c}\frac{P_i}{T_i},
\label{Equ:PropSec:GatwayLoad}
\end{equation}
where $g_c$ is the number of active gateways in the chiplet $c$, $P_i$ is the total number of transmitted packets during reconfiguration interval $i$, and $T_i$ is the duration of reconfiguration interval $i$ in cycles. Note that we assume a fixed packet size, otherwise $P_i$ should be the number of transmitted flits.  Moreover, we define a threshold for increasing and a threshold for decreasing the number of active gateways per chiplet. Accordingly, $T_{P_g}$ ($T_{N_g}$) is the threshold for increasing (decreasing) the number of gateways when the current number of active gateways is $g$. For $T_{P_g}$, we have:
\begin{equation}
T_{P_g}=T_{P_1}=T_{P_2}=T_{P_2}=....=T_{P_G}=L_m,
\label{Equ:PropSec:IncThreshold}
\end{equation}
where $G$ is the maximum number of active gateways per chiplet. When a gateway's load is higher than $L_m$, the gateway will suffer from notable congestion. Therefore, in such a case, the total number of active gateways on the chiplet should be increased to reduce the load on the congested gateway(s). As a result, $T_{P_g}$ is equal to $L_m$ in \eqref{Equ:PropSec:IncThreshold}. On the other hand, $T_{N_g}$ can be defined as:
\begin{equation}
T_{N_g}=L_m(1-\frac{1}{g}).
\label{Equ:PropSec:DecThreshold}
\end{equation}
To understand the rationale behind \eqref{Equ:PropSec:DecThreshold}, let us assume that we gradually reduce load $L$ from $L_m$ and try to find $T_{N_g}$. We need to reduce the number of active gateways from $g$ to $g-1$ when $L$ is small enough to avoid extra load on $g-1$ gateways (in the next reconfiguration interval). We can define this load reduction as: 
\begin{equation}
L_d=L_m -L_c,
\label{Equ:PropSec:LD}
\end{equation}
where $L_c$ is the current average gateway load of the chiplet. The sum of the reduced load of the $g$ active gateways is then: 
\begin{equation}
Sum(L_d)=L_d\times g
\label{Equ:PropSec:SL}.
\end{equation}
When $Sum(L_d)$ is equal to the maximum load of one gateway ($Sum(L_d)=L_m$), we can deactivate one gateway. Thus, we have:
\begin{equation}
\begin{array}{ c l }
T_{N_g}=L_c,   & \text{if~~~} Sum(L_d)=L_m. 
\end{array}
\label{Equ:PropSec:TNG}
\end{equation}
From \eqref{Equ:PropSec:LD}, \eqref{Equ:PropSec:SL}, and \eqref{Equ:PropSec:TNG}, we can calculate the threshold for decreasing the number of gateways: $T_{N_g}=L_m-L_d=L_m-\frac{L_m}{g}= L_m(1-\frac{1}{g})$.

\begin{figure}[t]
\centering
\includegraphics[scale=0.45]{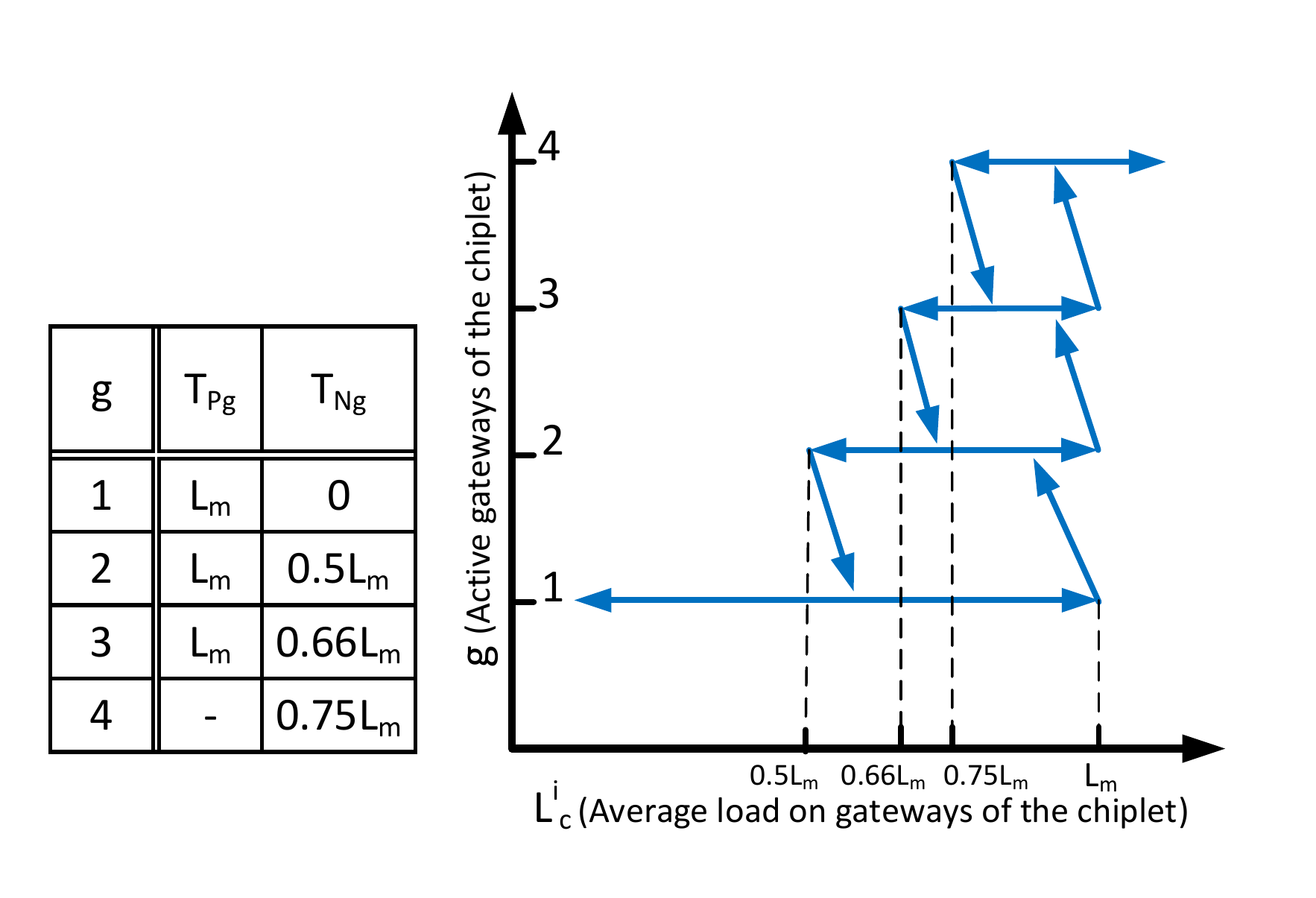}
\vspace{-0.17in}
\caption{Number of active gateways based on load changes.}
\label{Fig:PropSec:NumOfGates}
\vspace{-0.20in}
\end{figure}

As an example, the procedure to increase and decrease the number of active gateways ($g$) for a network with four gateways per chiplet is illustrated in Fig.~\ref{Fig:PropSec:NumOfGates}. Based on \eqref{Equ:PropSec:IncThreshold}, in each $g$ in the figure, if $L^i_c$ exceeds $L_m$, a new gateway will be activated ($g\rightarrow g+1$). On the other hand, according to \eqref{Equ:PropSec:DecThreshold}, if $L^i_c$ goes below $L_m(1-\frac{1}{g})$, one gateway will be deactivated ($g\rightarrow g-1$). Based on \eqref{Equ:PropSec:DecThreshold}, $T_{N_g}$ for different $g$ values is shown in a table in Fig.~\ref{Fig:PropSec:NumOfGates}. 

\begin{figure}[t]
\centering
\includegraphics[scale=0.75]{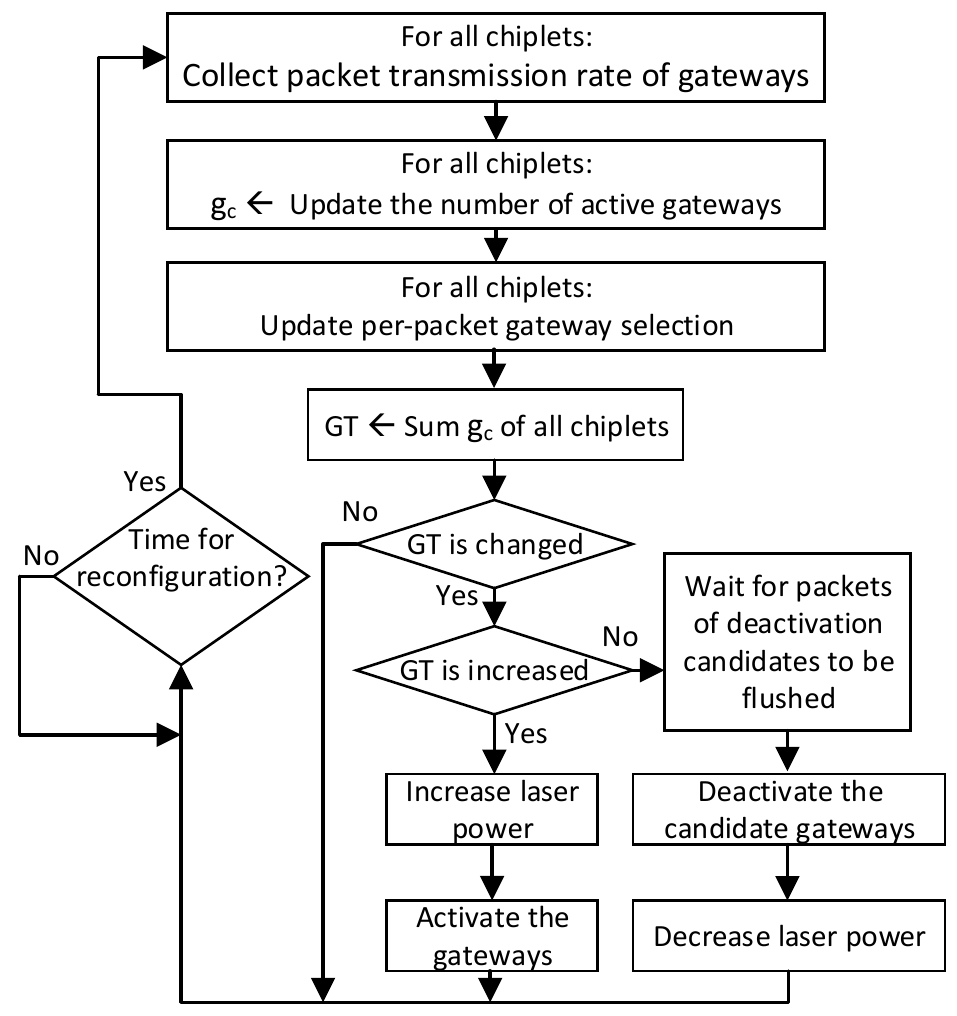}
\vspace{-0.15in}
\caption{Dynamic gateway management in ReSiPI.}
\label{Fig:PropSec:PowerLaserAlg}
\vspace{-0.25in}
\end{figure}

The dynamic gateway management algorithm in ReSiPI is illustrated in Fig.~\ref{Fig:PropSec:PowerLaserAlg}. The first step is to update the number of active gateways for each chiplet ($g_c$), which is initially set to the maximum allowed (four in our experiments in Section~\ref{Sec:Results}). After finding $g_c$, ReSiPI decides whether to change the number of active gateways, based on the procedure outlined above. If ReSiPI decides to increase the total number of active gateways ($GT$), first the laser power will be increased appropriately and then the additional gateways will be activated. On the other hand, to reduce the total number of active gateways, after waiting for packets of the candidate gateways to be routed (flushed), the gateways will be deactivated. After the gateway deactivation, the laser power can be reduced using a tunable SOA-based laser \cite{thakkar2016run}. 

We define a reconfiguration interval (i.e., epoch) at which we trigger the procedure to update the number of active gateways. A short reconfiguration interval will result in more frequent and responsive adaptation to traffic dynamics, while a long reconfiguration interval will result in a low reconfiguration overhead cost but also low responsiveness. We consider a reconfiguration interval length such that the update cost is negligible and ReSiPI is also able to efficiently adapt to traffic dynamics. The reconfiguration interval length that we consider (one million cycles) is significantly larger than the time to perform the reconfiguration/update processes, as will be further discussed in Section~\ref{Sec:Results}.

\subsection{Per-packet gateway selection}
For inter-chiplet packets, where routing over the photonic interposer network is required, a gateway in the source chiplet and a gateway in the destination chiplet are selected to perform the packet routing. Therefore, the routing process of an inter-chiplet packet is performed in three steps: 1) routing from the source router to the selected gateway on the source chiplet, 2) routing from the selected gateway on the source chiplet to the selected gateway on the destination chiplet, and 3) routing from the gateway on the destination chiplet to the destination router.

\begin{figure}[t]
\centering
\includegraphics[scale=0.39]{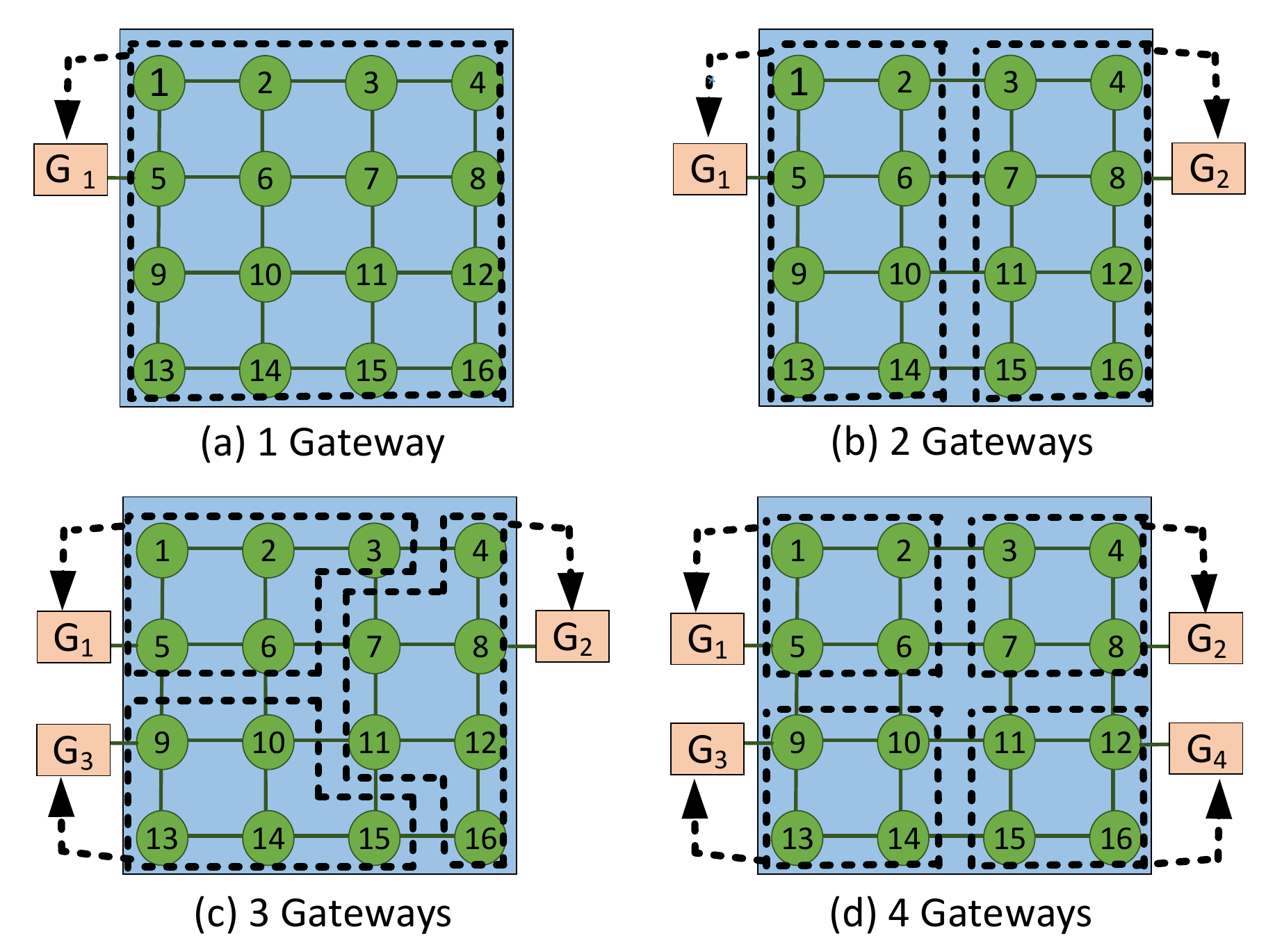}
\vspace{-0.2in}
\caption{An example of the adaptive gateway selection in ReSiPI for different number of activated gateways. The dashed boxes show the routers that will use a specific gateway (G).}
\label{Fig:PropSec:Selection}
\vspace{-0.20in}
\end{figure}

The gateway selection for each packet flow impacts the network performance, because it defines the assigned load on the gateways. An imbalanced gateway selection by packets can impose congestion on the gateways and degrade the overall performance \cite{taheri2021deft, taheri2021adele}. ReSiPI uses a dynamic per-packet gateway selection approach based on the number of active gateways in the source and destination chiplets. We take into account 1) gateways' traffic load and 2) router to gateway hop-counts in the analysis for gateway selection. To distribute the traffic load on the gateways, we try to balance the load on the gateways. Therefore, the average number of routers which can utilize the same gateway is $R_g=\frac{R}{g_c}$, where $R$ is the total number of routers on the chiplet. Then, we assign $R_g$ routers to a gateway in its vicinity. An example of gateway selection is shown in Fig.~\ref{Fig:PropSec:Selection}. In Fig.~\ref{Fig:PropSec:Selection}.a, only one gateway is activated, so all routers utilize this gateway. In Fig.~\ref{Fig:PropSec:Selection}.b, as there are two activated gateways, half of the routers ($R_g=8$) utilize the same gateway. For Figs.~\ref{Fig:PropSec:Selection}.c--d, similarly, the selection is done to balance the load on the gateways while each router is assigned to a gateway in its vicinity. For selecting the gateway at the destination chiplet, different gateway-selection scenarios are pre-analysed during design-time and the data related to the optimal destination gateway to minimize latency (for different scenarios of activated gateways at the destination chiplet) is stored in the gateway routers. 

Thus, for any packet being transmitted, the first routing step is performed in the source router based on the number of local active gateways, while the second step is performed in the source gateway based on the number of active gateways in the destination chiplet. In this way, the global information about active gateways only needs to be stored at gateways. Therefore, the source router is only aware of the number of active gateways in the source chiplet. On the other hand, the source gateway is aware of the number of active gateways in the destination chiplet. Design-time analysis helps to achieve a low-cost destination gateway selection that minimizes the latency to the destination router in the third step. This analysis utilizes hop count (from the destination gateway to the destination router) and number of active gateways in the destination chiplet to store selection decisions at the source gateway router, and these decisions are updated at every reconfiguration interval. 
 
\vspace{-0.025in}
\subsection{Reconfiguration controller architecture}
\begin{figure}[t]
\centering
\includegraphics[scale=0.88]{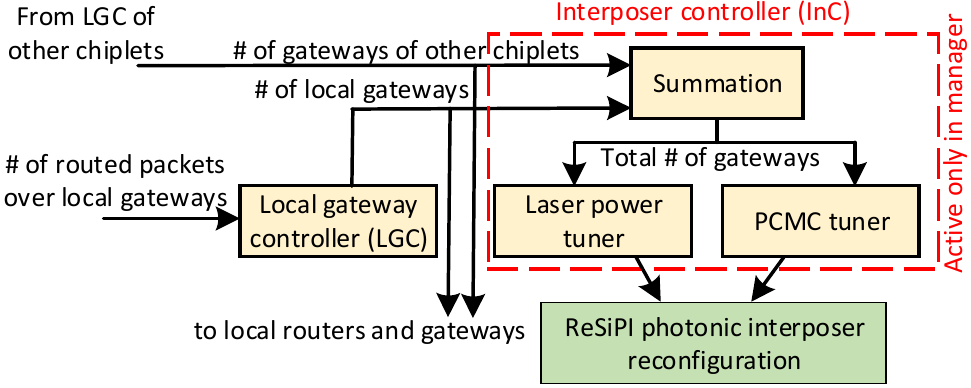}
\vspace{-0.3in}
\caption{ReSiPI's reconfiguration controller architecture.}
\label{Fig:PropSec:Controller}
\vspace{-0.15in}
\end{figure}

ReSiPI utilizes a controller in each chiplet to manage the gateway activation/deactivation, PCMCs, and laser power at the start of each reconfiguration interval. One of these controllers acts as the global manager that interacts with a local gateway controller (LGC) in each chiplet. The structure of ReSiPI's controller is shown in Fig.~\ref{Fig:PropSec:Controller}. All controllers in the system have this architecture but note that the interposer controller (InC) is only present in the global manager controller. LGCs decide on the number of active gateways on a chiplet, based on the number of routed packets over the chiplets active gateways. The LGC of each chiplet sends its number of active gateways to the interposer controller (InC) of the global manager controller (in one of the chiplets) at the end of a reconfiguration interval. InC sums the number of active gateways of chiplets ($g_c$) to define the total number of active gateways ($GT$) and tunes the PCMCs (based on \eqref{Equ:PropSec:CRActive}) and the laser power, as discussed in the earlier subsections. The controller overhead is discussed in Section \ref{Sec:Results}.  

\vspace{-0.05in}
\section{Simulation Results and Analysis}
\label{Sec:Results}
\begin{table}[t]
\caption{Simulation Setup.}\vspace{-0.1in}
\centering
\begin{tabular}{l l}
\hline
Parameter & value\\
\hline
Number of chiplets & 4 (each a 4$\times$4 mesh NoC)\\
Maximum gateways per chiplet & 1 for PROWAVES~\cite{narayan2020prowaves} \\
 & 4 for AWGR~\cite{fotouhi2019enabling} and ReSiPI\\
Gateways for memory controllers & 2\\
Gateway buffer size & 32 flits for PROWAVES\\
 & 8 flits for AWGR and  ReSiPI\\
Intra-chiplet router buffer size & 4 flits\\
Routing in chiplets & DeFT (deadlock free)~\cite{taheri2021deft}\\
Intra-chiplet NoC frequency & 1 Ghz\\
Data rate of optical link & 12 Gb/s per wavelength\\
Simulation cycles & 100~M (10~K for warm-up)\\
Reconfiguration interval duration & 1~M cycles\\
Packet size & 8 flits (each flit 32 bits)\\
\hline
\end{tabular}
\label{Table:ResultSec:SimSetup}
\vspace{-0.2in}
\end{table}

\vspace{-0.025in}
\subsection{Simulation setup}
To model 2.5D chiplet network platforms, we enhanced Noxim~\cite{catania2016cycle}, which is a cycle accurate NoC simulator. We used GEM5~\cite{binkert2011gem5} in full system mode to generate traffic traces of PARSEC benchmarks~\cite{bienia2011benchmarking}. We considered 64 x86 cores, where each core has a private L1 cache, four coherence directories, and four shared L2 cache
banks. We integrated the generated traffic traces into our enhanced Noxim simulator to analyze latency, power, and energy of the system. We compare ReSiPI with two photonic interposer networks, AWGR~\cite{fotouhi2019enabling} and PROWAVES ~\cite{narayan2020prowaves}. Our simulation configuration and setup is summarized in Table~\ref{Table:ResultSec:SimSetup}. We simulated a 2.5D network with four chiplets, where each chiplet has 16 cores connected by a 4$\times$4 mesh-based electronic NoC. We considered four gateways per chiplet where the gateways are connected to the chiplet similar to Fig.~\ref{Fig:PropSec:Selection}.d. The number of gateways per chiplet and the location of gateways are based on \cite{yin2018modular}. Unlike ReSiPI, PROWAVES advocates for changing the number of wavelengths to adapt a gateway's bandwidth to meet inter-chiplet bandwidth demands. We considered 16 wavelengths for PROWAVES, while ReSiPI uses 4 wavelengths. As a result, (number of wavelengths) $\times$ (number of gateways) in PROWAVES and ReSiPI are equal, to ensure that both have the same inter-chiplet bandwidth for a fair comparison. Moreover, we also considered the same buffer resource usage in both architectures. As ReSiPI has 4$\times$ gateways compared to PROWAVES, we considered 4$\times$ buffer size for PROWAVES (8 flit buffers in ReSiPI and 32 flit buffers in PROWAVES). AWGR~\cite{fotouhi2019enabling} requires one wavelength per gateway, so 18 wavelengths are used in the AWGR approach as the 2.5D network has 18 gateways in total. We used the silicon photonic power model in PROWAVES~\cite{narayan2020prowaves}. In the power model, laser power is 30 mW (per wavelength per waveguide), TIA power is 2 mW, thermal tuning power (per MR) is 3 mW, and driver power is 3 mW \cite{polster2016efficiency}.



\vspace{-0.05in}
\subsection{Design-space exploration: Optimal $L_m$}
\label{Sec:Results:DSE}
\begin{figure}[t]
\centering
\includegraphics[scale=0.6]{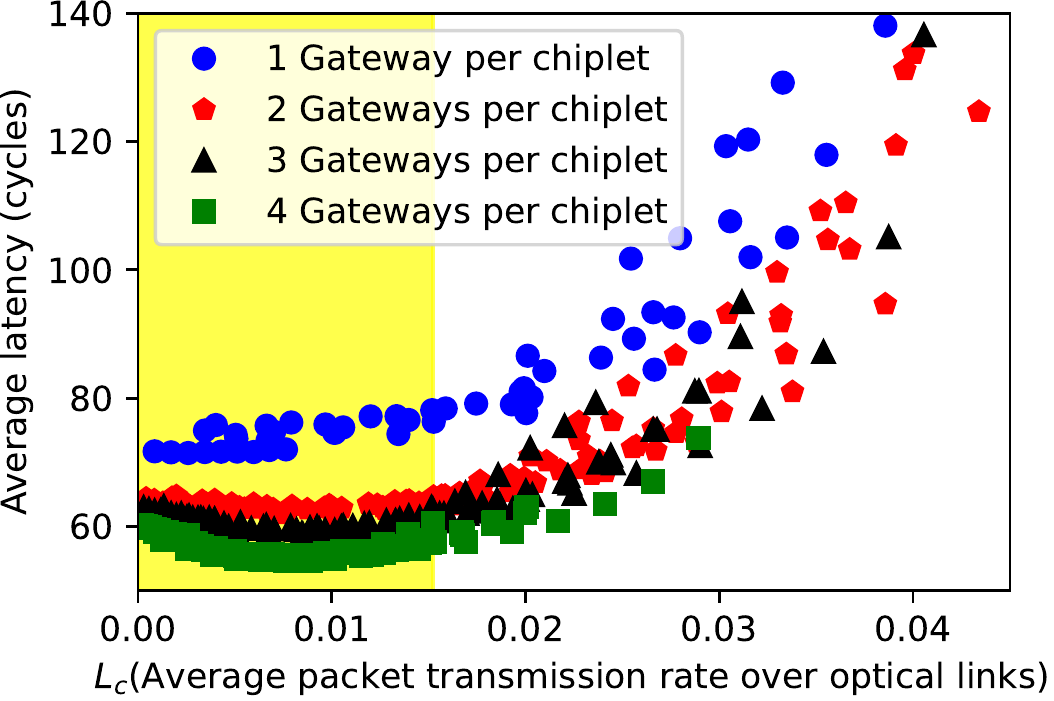}
\vspace{-0.15in}
\caption{Average latency vs. optical link transmission rate.}
\label{Fig:ResultSec:DSP}
\vspace{-0.25in}
\end{figure}

As discussed in Section~\ref{Sec:PropSec:NumOfGates}, $L_m$ is the maximum allowable load on a gateway. To find the optimal $L_m$, we evaluated our 2.5D network with various traffic and configuration scenarios. The results are shown in Fig.~\ref{Fig:ResultSec:DSP}. We simulated eight PARSEC applications: blackscholes, swaptions, streamcluster, facesim, fluidanimate, bodytrack, canneal, and dedup. The four different colors in Fig.~\ref{Fig:ResultSec:DSP} indicate the four main network configurations that we explored, with different numbers of gateways (1 to 4) per chiplet. Each simulation configuration, which corresponds to one point in the figure, gives us the average gateways' load $L_c$ (see \eqref{Equ:PropSec:GatwayLoad}) and the average packet latency. For the points with higher $L_c$, average latency is increased. Therefore, if we want to reduce the average latency, choosing a solution with lower $L_c$ is more efficient. However, a low $L_c$ means utilizing a larger number of active gateways, which will result in higher power consumption. This is because under the same traffic load, if the number of gateways is larger, less traffic load is assigned to each gateway. As a result, there is a trade-off in selecting $L_c$ with lower average latency or lower power consumption. In selecting $L_c$, we accept up to 10\% overhead in latency (empirically determined). The yellow-shaded region in Fig.~\ref{Fig:ResultSec:DSP} includes the points for which the average latency is smaller than 10\% overhead compared to the lowest average latency. Note that each point is compared with the points with the same number of active gateways. By accepting 10\% average latency overhead, $L_m$ is 0.0152 (maximum $L_c$ in the yellow-shaded region). With this value of $L_m$, the threshold for increasing the number of gateways ($T_{P_g}$) and the one for decreasing the number of gateways ($T_{N_g}$) can be calculated using \eqref{Equ:PropSec:IncThreshold} and \eqref{Equ:PropSec:DecThreshold}.

\begin{figure*}[t]
\centering
\includegraphics[scale=0.7]{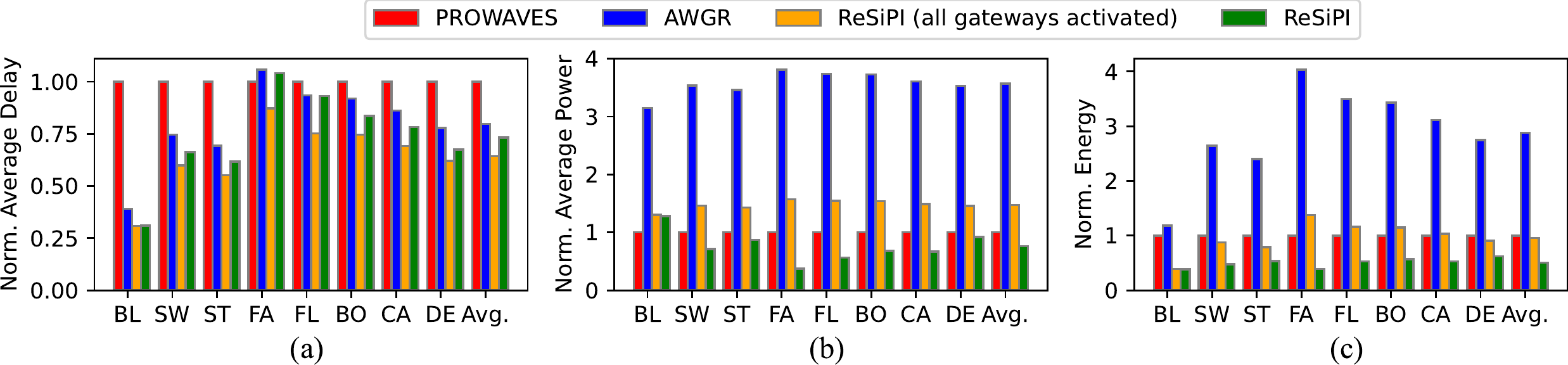}
\vspace{-0.15in}
\caption{(a) Normalized average latency, (b) Normalized average power, and (c) Normalized energy.}
\label{Fig:ResultSec:DelayPower}
\vspace{-0.2in}
\end{figure*}
\begin{figure}[t]
\centering
\includegraphics[scale=0.62]{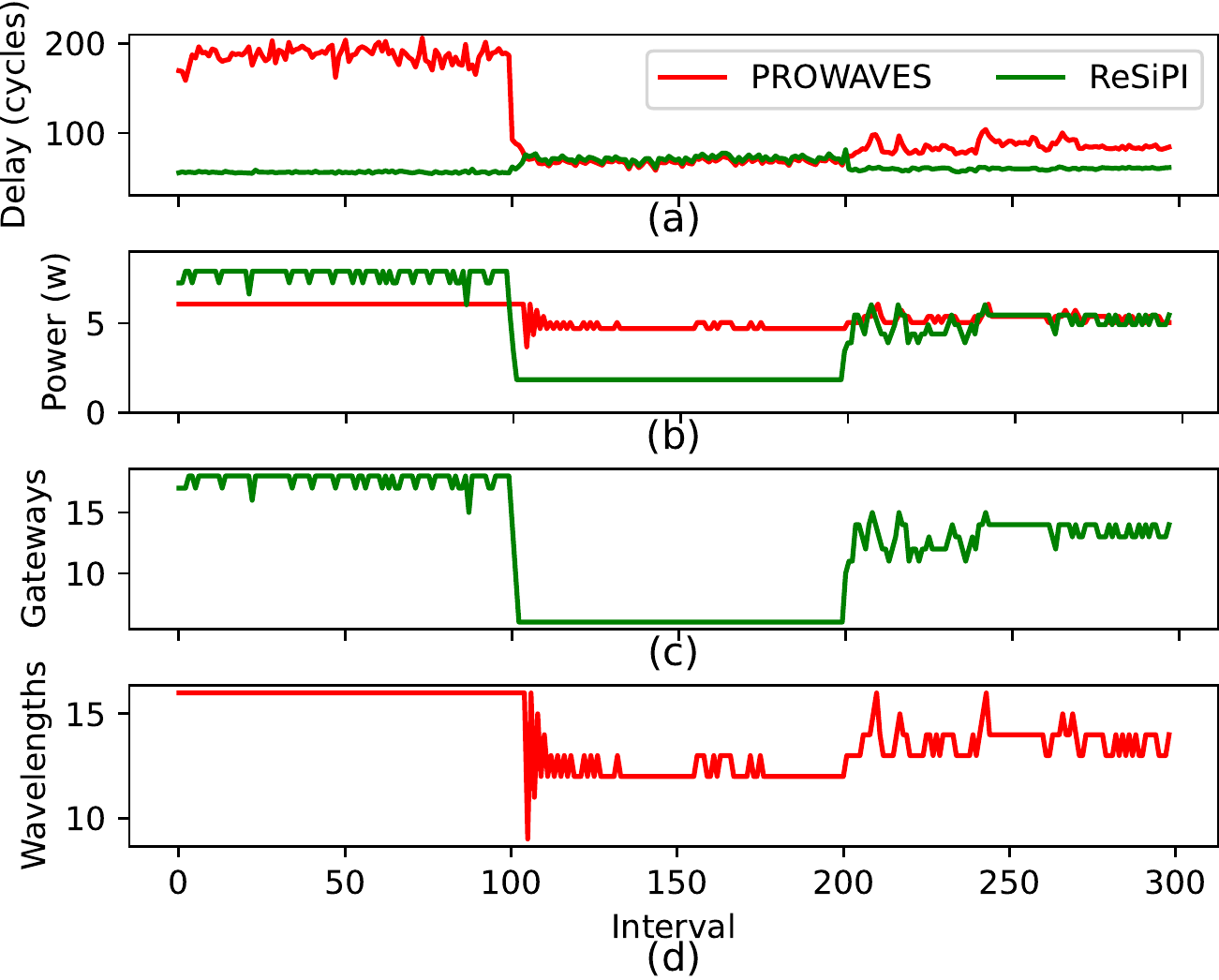}
\vspace{-0.15in}
\caption{Adaptivity comparison between ReSiPI and PROWAVES: (a) average delay, (b) average power, (c) number of activated gateways in ReSiPI, and (d) number of activated wavelengths in PROWAVES.}
\label{Fig:ResultSec:DynamicBehavior}
\vspace{-0.30in}
\end{figure}

\vspace{-0.05in}
\subsection{ReSiPI controller overhead}
We implemented ReSiPI's controller in HDL and synthesised it using Cadence Genus. We considered 1~Ghz clock frequency and 45~nm technology. The area and power overhead of the controller is summarized in Table.~\ref{hardware}. Both area and power are negligible compared to the budget of a chiplet (e.g., in \cite{narayan2020prowaves}, the chiplet area is 53.83~mm$^2$). In addition to the controller circuit, there are two important actions in the update process: 1) the reconfiguration time of PCMCs, and 2) the delay for tuning the laser power. We assumed the heater used in \cite{kato2017current}, to change PCMCs' state. According to \cite{kato2017current}, a PCM's state can be reconfigured in 100~ns. As our NoC frequency is 1~Ghz, the reconfiguration time of PCMCs is 100 cycles. We also assume an SOA-based laser and the time to tune the laser power is 20--50~ps \cite{thakkar2016run}. We consider a reconfiguration interval of one million cycles which is sufficient to capture major trends in traffic load changes, while it is quite large in comparison with the time to do the reconfigurations. The latency and power overhead for ReSiPI is considered in our simulation analysis in the rest of this section.

\begin{table}[t!]
\caption{Overhead analysis of ReSiPI's controller (see Fig.~\ref{Fig:PropSec:Controller}).}\vspace{-0.1in}
\centering
\begin{tabular}{|c||c|c|c|}
\hline
Parameter & LGC & InC & Total\\
\hline
\hline
Area ($\mu$m$^2$) & 314 & 104 & 418\\
\hline
Power ($\mu$W) &  172 & 787  & 959 \\
\hline
\end{tabular}
\label{hardware}
\vspace{-0.25in}
\end{table}

\vspace{-0.05in}
\subsection{Latency, power, and energy analysis}
The average latency, power, and energy results for all compared 2.5D network architectures are shown in Fig.~\ref{Fig:ResultSec:DelayPower}. The first two letters of each application is shown on the x-axis in the figure. In addition to ReSiPI, AWGR~\cite{fotouhi2019enabling}, and  PROWAVES~\cite{narayan2020prowaves}, we compared a variant of ReSiPI where all gateways are activated, to analyze the impact of dynamic inter-chiplet bandwidth management.

As shown in Fig.~\ref{Fig:ResultSec:DelayPower}.a, ReSiPI significantly improves the average latency in all the eight applications. On average, ReSiPI offers 37\% lower average latency due to its efficient architecture and bandwidth management. Moreover, as shown in Fig.~\ref{Fig:ResultSec:DelayPower}.b, ReSiPI consumes 25\% less power in comparison with PROWAVES, which is due to two main reasons. First, ReSiPI can handle inter-chiplet traffic with lower bandwidth budget as bisection bandwidth between chipets and the interposer is more distributed across the chiplets. Second, utilizing the PCM-based couplers, ReSiPI intelligently power-gates some part of the photonic interposer and saves laser power consumption. The AWGR approach \cite{fotouhi2019enabling} has high power consumption because 1) one wavelength is required for each AWGR port (gateway) and 2) AWGR's optical loss is high (1.8 dB loss based on \cite{fotouhi2019enabling}). Energy analysis is also shown in Fig.~\ref{Fig:ResultSec:DelayPower}.c, where ReSiPI offers a remarkable reduction across all the applications.

Fig.~\ref{Fig:ResultSec:DelayPower}.a shows that ReSiPI imposes a small average latency overhead compared to the ReSiPI variant with all gateways activated. This is because ReSiPI intelligently accepts a small latency overhead to considerably save on the power consumption, as shown in Fig.~\ref{Fig:ResultSec:DelayPower}.b. As we discussed in Section \ref{Sec:Results:DSE}, we chose $L_m$ while accepting 10\% overhead in the average latency to save on the power consumption in the design trade-off. Selecting a smaller $L_m$ slightly improves the average latency while imposing high power consumption overhead. Therefore, compared to when all the gateways are activated, ReSiPI greatly minimizes energy, as shown in Fig.~\ref{Fig:ResultSec:DelayPower}.c.

\vspace{-0.05in}
\subsection{Adaptivity analysis}
To contrast the adaptive behavior in ReSiPI and the best performing 2.5D network from prior work, PROWAVES (e.g., when traffic load changes), we simulated three applications in a sequence. Each application was executed for 100 million cycles (100 intervals). We measured latency and power of each reconfiguration interval to observe the adaptation behavior with ReSiPI and PROWAVES across reconfiguration intervals. For this analysis, we selected the applications with the highest load: Blackscholes, the lowest load: Facesim, and the median load: Dedup, respectively. Fig.~\ref{Fig:ResultSec:DynamicBehavior} shows the performance of ReSiPI and PROWAVES in terms of the average delay and the average power during the reconfiguration intervals. For the first 100 reconfiguration intervals, when Blackscholes is executing, which is the application with the highest load, ReSiPI can handle the traffic load and offers a low average latency. As shown in Fig~\ref{Fig:ResultSec:DynamicBehavior}.c, ReSiPI activates the maximum number of gateways (4$\times$4+2$=$18) in most of the cases to handle the traffic load with a small power overhead. During the Blackscholes application, although PROWAVES runs at its maximum bandwidth capacity with the maximum number of wavelengths (see Fig.~\ref{Fig:ResultSec:DynamicBehavior}.d), it is unable to adequately handle the traffic because the bandwidth is increased on the single gateway on each chiplet, rather than in a distributed manner across gateways in ReSiPI. Switching from Blackscholes to Facesim, ReSiPI adapts to the new traffic within three reconfiguration intervals only, whereas PROWAVES is unstable for five reconfiguration intervals. During the execution of Facesim, ReSiPI switches to a smaller number of active gateways and significantly reduces power consumption. ReSiPI imposes a small average latency overhead when executing Facesim. This is because ReSiPI finds the traffic load low and deactivates some unnecessary gateways. For the third application (Dedup), ReSiPI is again able to efficiently adapt to the traffic and manage the number of active gateways to achieve low power consumption. We also use the Dedup traffic to show the bandwidth distribution of ReSiPI next. \vspace{-0.15in}
\begin{figure}[t]
\centering
\includegraphics[scale=0.6]{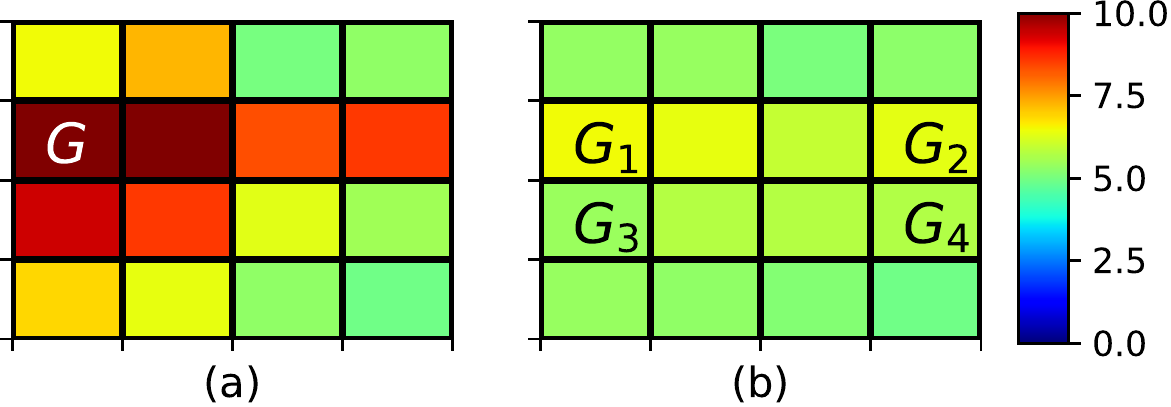}
\vspace{-0.15in}
\caption{Average residency of flits on the routers in the first chiplet in (a) PROWAVES and (b) ReSiPI.}
\label{Fig:ResultSec:Residency}
\vspace{-0.25in}
\end{figure}

\vspace{-0.05in}
\subsection{Bandwidth distribution analysis}
\label{Sec:Results:ChipletTraffic}
To further explain the performance differences between PROWAVES and ReSiPI, we monitored the residency of flits, which is the average time (in cycles) that flits stay in the router for both architectures.  Fig.~\ref{Fig:ResultSec:Residency} shows the average residency of one of the chiplets when using PROWAVES and ReSiPI. We do not show all the chiplets as the trend is similar in other chiplets. Although PROWAVES increases the bandwidth of gateways by increasing wavelengths, there is high congestion on the router connected to the gateway as shown in Fig.~\ref{Fig:ResultSec:Residency}.a (router $G$ in the figure). Moreover, the high congestion on the routers leads to back-pressure in the entire chiplet, creating high network congestion. On the other hand, as the load is more efficiently distributed among different routers in ReSiPI, the average residency of routers is low (see Fig.~\ref{Fig:ResultSec:Residency}.b). In ReSiPI, two gateways are often activated, which are connected to the routers at $G_1$ and $G_2$ in Fig.~\ref{Fig:ResultSec:Residency}.b. The distributed bandwidth enhancement in ReSiPI thus significantly improves network congestion over PROWAVES.

\vspace{-0.05in}
\section{Conclusion}
This paper presented ReSiPI which is a PCM-based reconfigurable silicon-photonic interposer network architecture for improving energy-efficiency in 2.5D chiplet systems. ReSiPI monitors the traffic load on the interposer and dynamically activates/deactivates gateways in the network. Activation of a larger number of gateways improves the average latency, while increasing the power consumption, and vice versa. ReSiPI's controller intelligently manages this latency-power trade-off and, therefore, can achieve 53\% improvement in network energy, in comparison with the best state-of-the-art 2.5D photonic  network. Results with real application traffic indicate that ReSiPI is a promising solution for an energy-efficient interposer network in emerging 2.5D chiplet platforms. 

\vspace{-0.05in}
\section*{Acknowledgment}
This work was supported by the National Science Foundation (NSF) under grant number CNS-2046226 and CCF-1813370.

\vspace{-0.05in}
\bibliographystyle{ACM-Reference-Format}
\bibliography{ref}

\end{document}